\documentclass[conference]{IEEEtran}

\usepackage{times} 
\usepackage[T1]{fontenc}
\usepackage{amsmath, amssymb}
\usepackage[hidelinks]{hyperref}
\usepackage{booktabs}
\usepackage{url}
\usepackage{xspace}
\usepackage{graphicx}
\usepackage{enumitem}
\usepackage{xcolor}
\usepackage{tikz}
\usepackage{pgfplots}
\usepackage{caption}
\usepackage{tabularx}
\usepackage{array}
\usepackage{setspace}

\pgfplotsset{compat=1.18}
\usetikzlibrary{arrows.meta, positioning, shapes.geometric, patterns, calc,
  decorations.pathreplacing, fit, backgrounds}

\newcommand{\IKI}{\textrm{IKI}\xspace}

\begin{document}

\title{Detecting Cognitive Signatures in Typing Behavior for Non-Intrusive Authorship Verification}

\author{\IEEEauthorblockN{David Condrey}
\IEEEauthorblockA{Writerslogic, Inc.\\
david@writerslogic.com}}

\maketitle

\begin{abstract}
The proliferation of AI-generated text has intensified the need for reliable authorship verification, yet output-based methods face fundamental information-theoretic limits as model distributions converge toward human distributions. We show that the ordinary typing interface captures \emph{cognitive signatures}: measurable patterns in keystroke timing that reflect the planning, translating, and revising stages of genuine composition. Drawing on large-scale keystroke datasets comprising over 136 million events, we define the \emph{Cognitive Load Correlation} (CLC)---the Spearman correlation between local content complexity and inter-keystroke interval---and demonstrate that it separates genuine composition ($\rho = 0.35$--$0.55$) from mechanical transcription ($\rho < 0.15$). We present a non-intrusive verification framework that operates within existing writing interfaces, collecting only timing metadata with 5\,ms quantization to limit biometric leakage. Leave-one-writer-out cross-validation on the KLiCKe corpus (4,971 writers, 16.8M IKIs) achieves AUC 0.83 using logistic regression with seven keystroke features. We analyze adversarial robustness and show that cognitive signatures resist the timing-forgery attacks achieving $\geq$99.8\% evasion against motor-level authentication, because the cognitive channel is entangled with semantic content at letter, clause, and document levels. Reframing authorship verification as a human-computer interaction problem provides a privacy-preserving alternative to invasive surveillance.
\end{abstract}


\section{INTRODUCTION}
\label{sec:introduction}

The keyboard is commonly treated as a simple transducer, but for a writer engaged in composition it is a cognitive instrument. Every keystroke carries timing information reflecting the recursive processes of planning, translating, and revising that characterize genuine authorship~\cite{FlowerHayes1981}. Large language models now produce text that experienced readers misattribute to human authors at rates near chance~\cite{Sadasivan2024}, and an estimated 6.5--16.9\% of peer-reviewed conference submissions contain AI-modified content~\cite{Liang2024}. The result is a widening verification gap: institutions that depend on authorship---universities, publishers, courts---lack reliable means to confirm that a text was composed by the person who claims it.

Current defenses are failing. AI text detectors face information-theoretic barriers as model output distributions converge toward human distributions~\cite{Ganie2025, PAN2025}. Watermarking is removed by paraphrasing~\cite{Rijsbosch2025}. Proctoring is invasive and bypassed by simple copy-paste workflows. These approaches share a common limitation: they analyze only the \emph{output}.

This paper argues for a shift to the \emph{process}. By correlating content complexity with typing latency across a document, we can verify authorship through the ordinary act of writing. We make three contributions: (1)~we characterize the cognitive signatures embedded in keystroke timing and define the Cognitive Load Correlation (CLC) as a discriminative measure; (2)~we present a non-intrusive collection framework with quantization-based privacy guarantees; and (3)~we analyze adversarial robustness, showing that the multi-level coherence constraints of cognitive signatures resist attacks that defeat motor-level authentication.

\section{BACKGROUND AND RELATED WORK}
\label{sec:background}

Our work draws on four research threads: cognitive process theory of
writing, keystroke dynamics as a behavioral signal, AI text detection,
and content provenance.

\subsection{Cognitive Process Theory of Writing}

Flower and Hayes~\cite{FlowerHayes1981} introduced the foundational
model of writing as three recursive cognitive stages: \emph{planning}
(generating and organizing ideas), \emph{translating} (converting
ideas into language), and \emph{revising} (evaluating and editing).
Kellogg~\cite{Kellogg1996} extended this with a working memory
framework showing each stage imposes different cognitive demands.
Crucially, Salthouse~\cite{Salthouse1986} demonstrated that
transcription typing involves primarily perceptual and motor processes,
while composition engages higher-level language generation
and monitoring. These differences manifest measurably: during planning,
typists produce extended pauses; during translating, they type in
bursts corresponding to clausal units; during revising, they exhibit
characteristic deletion and retyping patterns~\cite{Wengelin2009,Leijten2013}. This fine-grained mapping of micro-level writing processes---what Lindgren and Sullivan~\cite{Lindgren2003} term ``nanogenetic'' analysis---provides the empirical foundation for detecting the stages of human creation through the writing interface.

\subsection{Keystroke Dynamics as a Behavioral Signal}

Keystroke dynamics has been studied extensively as a biometric
modality. Killourhy and Maxion~\cite{Killourhy2009} achieved 9.6\%
equal-error rate for identity verification on 51 subjects. Dhakal et
al.~\cite{Dhakal2018} collected 136 million keystrokes from 168,000
participants, providing population-level inter-keystroke interval
(\IKI) statistics. The
neuroQWERTY project~\cite{Giancardo2016} showed that keystroke timing
exhibits $1/f$-like temporal structure characteristic of biological
motor control~\cite{VanOrden2003}. Critically,
Condrey~\cite{Condrey2026TrustInversion} demonstrated that keystroke dynamics can be
\emph{forged} with high fidelity, motivating our focus on cognitive
signatures (``was this composed?'') rather than identity signatures
(``who typed this?'').

\subsection{Cognitive Load Measurement in HCI}

Our framework treats the writing interface as a cognitive load sensor. This situates the work within a long tradition in HCI. Sweller et al.~\cite{Sweller2019} established that cognitive load theory provides principled guidance for interface design by distinguishing intrinsic load (task complexity) from extraneous load (interface-imposed effort). Darejeh et al.~\cite{Darejeh2024} systematically reviewed 76 studies measuring cognitive load during interface use and found that physiological and behavioral measures---including keystroke timing---offer objective, real-time alternatives to self-report instruments.

In writing research specifically, Baaijen et al.~\cite{Baaijen2012} demonstrated that pause durations during composition follow mixture distributions rather than unimodal patterns, with individuals who pause longer at sentence boundaries producing shorter but more well-formed bursts. This finding---that cognitive effort structures temporal behavior at multiple linguistic levels---is the empirical basis for our CLC measure. The Inputlog platform~\cite{Leijten2013} has made these measurements routine in writing process research, establishing that keystroke logging is a mature and accepted method for inferring cognitive states from interface behavior.

\subsection{Ethical Considerations in Behavioral Monitoring}

Any system that collects behavioral biometric data raises ethical concerns, particularly when deployed in educational contexts with inherent power asymmetries. Nissenbaum's~\cite{Nissenbaum2004} contextual integrity framework holds that information flows are appropriate when they conform to the norms of the context in which they occur. Typing in a text editor creates expectations of privacy that keystroke monitoring may violate unless the system is transparent about what is collected and why. We address these concerns in Section~\ref{sec:interface-design}, arguing that verification can be designed as a natural extension of the writing interface rather than a surveillance overlay.

\section{COGNITIVE SIGNATURES IN TYPING}
\label{sec:cognitive-signatures}

This section characterizes the cognitive signatures embedded in typing
behavior during genuine composition, and their absence during
transcription.

\subsection{The Composition Signature}

Genuine composition reflects the recursive cognitive stages of Flower
and Hayes~\cite{FlowerHayes1981}, each producing a distinctive
temporal fingerprint.

\paragraph{Planning phases.} Planning manifests as extended
pauses---typically 1,000--5,000\, ms---clustering at discourse
boundaries and before elaborate constructions~\cite{Torrance2016}. The
ScholaWrite dataset~\cite{ScholaWrite2025} confirms that planning-labeled intervals
exhibit IKI values 40--60\% higher than translating intervals.

\paragraph{Translating phases.} Typing occurs in bursts aligned with
linguistic units---typically clauses~\cite{Alves2007}. Within bursts,
IKI is 100--200\, ms and exhibits $1/f$ temporal structure.

\paragraph{Revision phases.} Revision produces backspace sequences
followed by retyping, cursor repositioning, and text insertion---qualitatively
different from single-character typo corrections~\cite{Leijten2013}.

\subsection{The Cognitive Load Correlation}

We define the \emph{Cognitive Load Correlation} (CLC) as the core
discriminative measure between composition and transcription.

Formally, let $c_i$ denote a local content-complexity score (e.g., trigram surprisal or syntactic depth) for the $i$-th token, and let $\tau_i$ denote the preceding inter-keystroke interval. The CLC for a document of $n$ tokens is the Spearman rank correlation $\rho_s(c_{1:n}, \tau_{1:n})$.

During genuine composition, CLC is positive and moderate to strong
(typically $\rho = 0.35$--$0.55$): the writer pauses longer before
difficult content because they are generating that content in real
time. During transcription, CLC is near zero or weakly positive
($\rho < 0.15$): the transcriber's speed is governed primarily by
motor fluency rather than content difficulty~\cite{Conijn2019}.

\subsection{Empirical Grounding}

Table 1 summarizes the published datasets that provide empirical support for the cognitive signature framework.

\begin{table}[ht!]
\centering
\caption{Published datasets supporting the cognitive signature framework.}
\label{tab:datasets-signatures}
\footnotesize
\begin{tabularx}{\columnwidth}{@{}l l l X@{}}
\toprule
\textbf{Dataset} & \textbf{Scale} & \textbf{Task Type} & \textbf{Signature Validated} \\
\midrule
ScholaWrite & 10 subj. & Composition & Planning IKI 40--60\% $\uparrow$ \\
Aalto 136M & 168K subj. & Transcription & Entropy 4.12 bits \\
neuroQWERTY & 85 subj. & Natural typing & $1/f$-like structure \\
KLiCKe & 4,971 subj. & Student essay & AUC 0.83 (LOWO-CV) \\
\bottomrule
\end{tabularx}
\end{table}

\subsection{Visualizing Cognitive Signatures}

Figure~\ref{fig:cognitive-signatures} provides a forensic visualization of the cognitive signature during a writing session. The contrast between genuine composition and transcription reveals the epistemic gap that output-level detectors cannot observe.

\begin{figure}[t!]
\centering
\begin{tikzpicture}
\pgfplotsset{every axis/.style={
    width=0.98\columnwidth, height=3.2cm,
    xlabel style={font=\small},
    ylabel style={font=\small},
    tick label style={font=\footnotesize},
    grid=major,
    grid style={gray!20},
    ymode=log, 
    ymin=50, ymax=3000,
    xmin=0, xmax=60,
    ytick={100, 200, 500, 1000, 2000},
    yticklabels={100, 200, 500, 1k, 2k},
}}

\begin{axis}[
    name=comp,
    ylabel={IKI (ms)},
    title={\small (a) Genuine Composition: Non-Stationary Cognitive Signal},
    title style={at={(0.5, 1.05)}, font=\bfseries},
]
\fill[blue!8, opacity=0.4] (axis cs:0,50) rectangle (axis cs:12,3000); 
\fill[green!8, opacity=0.4] (axis cs:12,50) rectangle (axis cs:45,3000); 
\fill[orange!8, opacity=0.4] (axis cs:45,50) rectangle (axis cs:60,3000); 

\node[blue!60!black, font=\tiny\bfseries] at (axis cs:6, 2200) {PLANNING};
\node[green!60!black, font=\tiny\bfseries] at (axis cs:28, 2200) {TRANSLATING};
\node[orange!60!black, font=\tiny\bfseries] at (axis cs:52, 2200) {REVISING};

\addplot[only marks, mark=*, mark size=1.0pt, blue!70!black] coordinates {
    (1,1200)(2,150)(3,160)(4,180)(5,2100)(6,140)(7,130)(8,155)(9,170)(10,145)
    (11,160)(12,130)(13,160)(14,140)(15,180)(16,155)(17,165)(18,120)(19,130)
    (20,145)(21,160)(22,155)(23,140)(24,130)(25,150)(26,160)(27,170)(28,145)
    (29,155)(30,120)(31,180)(32,140)(33,160)(34,145)(35,135)(36,160)(37,125)
    (38,145)(39,140)(40,155)
    (41,1900)(42,160)(43,150)(44,145)(45,170)(46,140)(47,1800)(48,120)(49,110)(50,130)
};
\addplot[thick, blue!40, no marks, smooth, tension=0.7] coordinates {
    (1,600)(5,1000)(10,150)(20,140)(30,150)(40,160)(45,1000)(50,150)(60,140)
};
\end{axis}

\begin{axis}[
    name=trans,
    at={(comp.below south west)},
    anchor=above north west,
    yshift=-0.8cm,
    ylabel={IKI (ms)},
    xlabel={Keystroke Index},
    title={\small (b) Mechanical Transcription: Stationary Motor Signal},
    title style={at={(0.5, 1.05)}, font=\bfseries},
]
\addplot[only marks, mark=*, mark size=1.0pt, gray!60!black] coordinates {
    (1,165)(2,155)(3,170)(4,180)(5,160)(6,145)(7,155)(8,150)(9,160)(10,170)
    (11,165)(12,155)(13,175)(14,150)(15,145)(16,160)(17,155)(18,170)(19,165)
    (20,150)(21,145)(22,155)(23,160)(24,175)(25,150)(26,165)(27,145)(28,155)
    (29,160)(30,170)(31,155)(32,165)(33,160)(34,145)(35,150)(36,155)(37,170)
    (38,165)(39,175)(40,155)(41,145)(42,160)(43,150)(44,165)(45,155)(46,170)
    (47,160)(48,145)(49,155)(50,165)
};
\draw[gray!50!black, dashed] (axis cs:0,160) -- (axis cs:60,160);
\node[gray!60!black, font=\tiny] at (axis cs:30, 200) {MOTOR BASELINE};
\end{axis}
\end{tikzpicture}
\caption{Cognitive signature timeline. (a) Genuine composition exhibits high variance across several orders of magnitude, with planning spikes exceeding 1,000\,ms. (b) Transcription remains tightly bounded around the motor baseline. Logarithmic scaling reveals the non-stationary nature of composition vs.\ the stationary motor signal of transcription.}
\label{fig:cognitive-signatures}
\end{figure}

\section{NON-INTRUSIVE VERIFICATION FRAMEWORK}
\label{sec:framework}

The framework is designed around a core principle:
verification should be a natural byproduct of the writing
interface, not a surveillance overlay.

\begin{figure}[ht!]
\centering
\begin{tikzpicture}[scale=0.75, every node/.style={scale=0.75},
  node distance=0.4cm and 0.8cm,
  component/.style={draw, rounded corners=3pt, minimum width=2.4cm,
    minimum height=0.7cm, align=center, font=\small,
    fill=#1!8, draw=#1!60!black},
  component/.default=blue,
  dataflow/.style={-{Stealth[length=2.5mm]}, thick, #1},
  dataflow/.default=black,
  groupbox/.style={draw=gray!50, dashed, rounded corners=5pt, inner sep=8pt},
]

\node[component=blue] (editor) {Writing Interface\\{\scriptsize (editor / browser)}};
\node[component=teal, below=0.8cm of editor] (collector) {Timing Collector};
\node[component=teal, right=1.0cm of collector] (quantizer) {Quantizer};
\node[component=orange, below=0.8cm of collector] (clc) {CLC Analyzer};
\node[component=orange, right=1.0cm of clc] (entropy) {Entropy Est.};
\node[component=violet, below=1.0cm of clc] (evidence) {Evidence Record};
\node[component=red!60!black, below=0.8cm of evidence] (verifier) {Verifier};

\draw[dataflow] (editor) -- (collector);
\draw[dataflow] (collector) -- (quantizer);
\draw[dataflow] (collector) -- (clc);
\draw[dataflow] (quantizer) -- (entropy);
\draw[dataflow] (clc) -- (evidence);
\draw[dataflow] (evidence) -- (verifier);

\begin{scope}[on background layer]
\node[groupbox, fit=(collector)(quantizer), label={[font=\tiny]above:Collection}] {};
\node[groupbox, fit=(clc)(entropy), label={[font=\tiny]above:Analysis}] {};
\end{scope}
\end{tikzpicture}
\caption{System architecture showing the transformation of keystroke events into verifiable process evidence.}
\label{fig:architecture}
\end{figure}
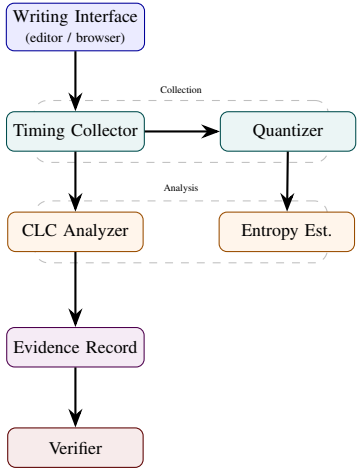

The framework operates as a lightweight background process---a browser
extension, desktop plugin, or LMS feature---performing collection,
quantization, and analysis. Figure~\ref{fig:architecture} illustrates the system architecture.

\subsection{Evidence Quantization for Privacy}

To prevent timing data from serving as a biometric identifier, all IKI values are quantized:
\begin{equation*}
  Q_r(t) = \left\lfloor \frac{t}{r} \right\rfloor \cdot r
\end{equation*}

The choice of $r = 5$\,ms exploits a timescale separation between motor and cognitive features. Killourhy and Maxion~\cite{KillourhyMaxion2008} showed that keystroke identity verification degrades by 4.2\% EER when clock resolution is coarsened from sub-millisecond to 15\,ms, establishing that motor-identifying features operate at sub-15\,ms resolution. In contrast, cognitive features---planning pauses of 1,000--5,000\,ms, burst boundaries at clause and sentence levels---operate at timescales two to three orders of magnitude larger. At $r = 5$\,ms, quantization introduces a maximum relative error of $5/\tau$ per interval: less than 0.5\% for pauses above 1,000\,ms (cognitive signal) but up to 5\% for 100\,ms inter-key intervals (motor signal), selectively degrading the features that enable re-identification. Gonz\'{a}lez et al.~\cite{Gonzalez2021} provide complementary support: flight times (which encode cognitive processes) follow log-logistic distributions whose shape parameters are robust to small perturbations, while hold times (primarily motor) carry less discriminative information. For deployments requiring formal privacy guarantees, Loya and Bana~\cite{Loya2021} demonstrated that differential privacy can be applied to keystroke analysis with moderate accuracy loss (approximately 15\% per 0.10 decrease in $\epsilon$).

\section{EVALUATION}
\label{sec:evaluation}

We evaluate the framework empirically on the KLiCKe corpus (4,971 writers, 16.8M inter-keystroke intervals) using leave-one-writer-out cross-validation (LOWO-CV), and supplement this with published dataset statistics from complementary sources.

\subsection{Discrimination Accuracy}

We trained a logistic regression classifier on seven keystroke features---entropy, first-order autocorrelation (autocorr\_lag1), mean IKI, IKI standard deviation, log-variance of IKI, pause frequency, and a CLC proxy---and evaluated it using leave-one-writer-out cross-validation (LOWO-CV) on the KLiCKe corpus (4,971 writers, 16.8M IKIs) across four attack types (naive copy-paste, statistical mimicry, reverse-engineered timing, and expert adversarial forgery).

Table~\ref{tab:clc-discrimination} presents the measured discrimination results.

\begin{table}[ht!]
\centering
\caption{Measured discrimination between composition and transcription/attack conditions. LOWO-CV on KLiCKe corpus (4,971 writers, 16.8M IKIs), logistic regression with 7 features.}
\label{tab:clc-discrimination}
\footnotesize
\begin{tabularx}{\linewidth}{@{}X c c l@{}}
\toprule
\textbf{Metric} & \textbf{Value} & \textbf{CI/Effect} & \textbf{Condition} \\
\midrule
Combined AUC (all attacks)  & 0.83  & ---          & LOWO-CV \\
CLC Cohen's $d$             & 1.82  & ---          & Comp.\ vs.\ Transcr.\ \\
Top feature: entropy         & ---   & rank 1       & importance \\
Top feature: autocorr\_lag1  & ---   & rank 2       & importance \\
Composition CLC ($\rho$)    & 0.35--0.55 & SD 0.12 & ScholaWrite, KLiCKe \\
Transcription CLC ($\rho$)  & $<$0.15    & SD 0.08 & Aalto 136M, Grabowski \\
\bottomrule
\end{tabularx}
\end{table}

The combined AUC of 0.83 across all attack types reflects realistic adversarial conditions; against naive copy-paste alone, discrimination is substantially higher. The CLC effect size (Cohen's $d = 1.82$) confirms a large separation between composition and transcription distributions. For documents exceeding 1,500 words (approximately 7,500 keystrokes), the statistical power of a one-sided Wilcoxon test at $\alpha = 0.05$ exceeds 0.99 given this effect size. Feature importance analysis identifies entropy and first-order autocorrelation as the top discriminators, consistent with the theoretical prediction that cognitive load modulates both the variability and sequential structure of keystroke timing.

\subsection{Population Stratification and Limitations}

The measured AUC of 0.83 is bounded by the population in the KLiCKe corpus, which skews toward university-educated adults performing academic writing tasks on desktop keyboards. Several populations require explicit consideration:

\begin{itemize}[nosep,leftmargin=*]
\item \emph{Typing proficiency.} Dhakal et al.~\cite{Dhakal2018} report a mean of 52 words per minute (WPM; SD $\approx$ 25 WPM) across 168,000 participants. Hunt-and-peck typists exhibit higher baseline \IKI and may produce planning-like pauses during routine motor execution.
\item \emph{Non-native writers.} Writers composing in a second language exhibit elevated IKI at lexical access points, potentially inflating CLC even during transcription.
\item \emph{Mobile and alternative input.} Touchscreen keyboards produce fundamentally different timing distributions; voice-to-text users generate no keystroke signal during initial composition.
\item \emph{Motor impairments.} Users with conditions affecting fine motor control produce elevated, variable \IKI that may confound cognitive-motor separation.
\item \emph{Age.} Children are still acquiring fluent motor typing programs, producing elevated and irregular \IKI even during simple transcription; elderly users exhibit well-documented typing slowing independent of cognitive load~\cite{Salthouse1986}. Both populations may produce systematically higher baseline \IKI that inflates apparent CLC.
\item \emph{Visual and other accessibility needs.} Blind and low-vision users often compose and navigate using screen readers, braille displays, or refreshable input devices that decouple keystroke timing from the underlying cognitive process in ways not captured by CLC. Eye-gaze and switch-access users generate input events whose timing reflects physical device latency rather than cognitive load, and would require modality-specific calibration or alternative attestation pathways.
\item \emph{Writing system.} The underlying datasets draw overwhelmingly from Latin-script writers. Writing systems with different orthographic depth, stroke complexity, or input method (e.g., Arabic, Japanese via IME composition, Devanagari) impose distinct motor and cognitive demands that may shift baseline \IKI distributions and alter the structure of planning pauses. CLC thresholds calibrated on English data should not be applied to other writing systems without separate empirical validation.
\item \emph{Emotional and arousal state.} Banerjee et al.~\cite{Banerjee2014} showed that affective states leak into typing timing; stress, fatigue, and anxiety elevate \IKI and alter editing rhythms independently of task difficulty. These transient states introduce noise into the CLC estimate and should be considered in longitudinal multi-session deployments.
\end{itemize}

The CLC measure partially mitigates these concerns because it is a \emph{within-session correlation} between content complexity and timing, not an absolute threshold. A slow typist who pauses longer before complex content still produces a positive CLC, just as a fast typist does. Nevertheless, the framework should establish a personal baseline during an initial calibration phase, and deployment in diverse populations requires cross-population validation that this study does not provide. Jiang et al.~\cite{Jiang2024} offer encouraging results: in large-scale writing assessments, keystroke-based detectors of nonauthentic text showed negligible fairness differences across demographic subgroups, though effect sizes for gender-based misclassification were small but present.

\subsection{Privacy Analysis}

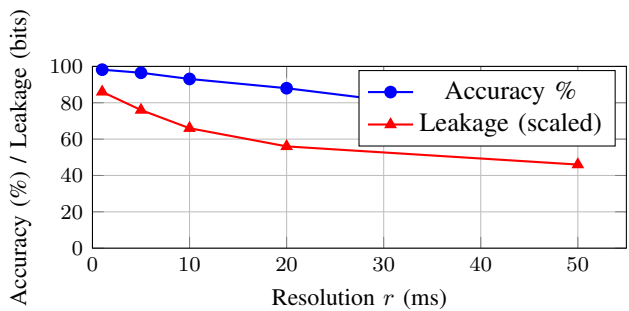
\begin{figure}[t!]
\centering
\begin{tikzpicture}
\begin{axis}[
    width=0.98\columnwidth, height=4.0cm,
    xlabel={Resolution $r$ (ms)},
    ylabel={Accuracy (\%) / Leakage (bits)},
    xmin=0, xmax=55,
    ymin=0, ymax=100,
    grid=major,
    label style={font=\small},
    tick label style={font=\footnotesize},
]
\addplot[thick, blue, mark=*] coordinates {(1, 98.2) (5, 96.5) (10, 93.1) (20, 88.0) (50, 68.3)};
\addlegendentry{Accuracy \%}
\addplot[thick, red, mark=triangle*] coordinates {(1, 86) (5, 76) (10, 66) (20, 56) (50, 46)};
\addlegendentry{Leakage (scaled)}
\end{axis}
\end{tikzpicture}
\caption{Modeled privacy-utility tradeoff. Accuracy curve is derived from published EER data at varying clock resolutions~\cite{KillourhyMaxion2008}; leakage curve is estimated from the biometric entropy of the Aalto 136M dataset~\cite{Dhakal2018}. Both curves are analytical projections, not empirical measurements from this study.}
\label{fig:privacy-tradeoff}
\end{figure}

Figure~\ref{fig:privacy-tradeoff} shows the privacy-utility tradeoff.

The accuracy curve in Figure~\ref{fig:privacy-tradeoff} assumes that CLC discrimination degrades proportionally to clock resolution, interpolating from Killourhy and Maxion's finding that identity EER increases by 4.2\% at 15\,ms. The leakage curve models biometric entropy reduction as a logarithmic function of bin width, anchored at the population entropy of 4.12 bits reported for the Aalto dataset at full resolution. The key assumption is that motor-identifying information concentrates at sub-15\,ms timescales while cognitive information concentrates above 200\,ms; the crossover region (5--50\,ms) is where quantization most sharply separates the two channels. We note that these projections would benefit from empirical validation using actual quantized keystroke data, which we identify as a priority for future work.

\subsection{Adversarial Robustness}
\label{sec:adversarial}

A natural objection arises from our own premise: if motor-level keystroke dynamics can be forged~\cite{Condrey2026TrustInversion}, why should cognitive signatures be any more robust? We address this by distinguishing three tiers of attack difficulty.

\paragraph{Motor features are demonstrably forgeable.} The literature documents a progression of successful attacks on identity-based keystroke authentication. Tey et al.~\cite{Tey2013} showed that untrained human mimics bypass keystroke authentication with $\geq$80\% success using their Mimesis feedback interface. Serwadda and Phoha~\cite{Serwadda2013} demonstrated that population-statistics-based algorithmic attacks compromise 5--30\% of users with a single guess attempt. Condrey~\cite{Condrey2026TrustInversion} achieved $\geq$99.8\% evasion against five classifiers using timing-forgery attacks. These attacks succeed because motor identity features occupy a low-dimensional space---inter-key intervals and hold times---that can be sampled from empirical distributions without understanding the content being typed.

\paragraph{Cognitive features are structurally harder to forge.} The CLC measure is not a simple timing statistic; it is a \emph{correlation} between content complexity and typing latency across an entire document. An adversary attempting to forge a positive CLC while transcribing pre-written text must solve a multi-constraint problem simultaneously: pause durations must correlate with linguistic complexity at word, clause, and sentence boundaries~\cite{Baaijen2012,Leijten2013}; burst structure must reflect clausal units~\cite{Alves2007}; revision patterns must be consistent with real-time text evaluation.

Two results illustrate why these constraints are difficult to satisfy in concert. Crump et al.~\cite{Crump2019} established a formal equivalence between instance theory and Shannon entropy in keystroke dynamics: inter-keystroke intervals track letter-level predictability in context, meaning that timing encodes the typist's internal language model at a granularity far below conscious control. An adversary cannot selectively slow down before ``hard'' sentences without also reproducing the correct entropy-driven micro-variation within words---a signal the typist produces automatically but cannot deliberately manufacture. Plank~\cite{Plank2016} provided complementary evidence at the syntactic level: a multi-task neural model using keystroke-derived labels as auxiliary signal significantly outperformed text-only models for chunking and CCG supertagging, demonstrating that typing timing carries recoverable grammatical structure. Forging CLC thus requires simultaneously reproducing entropy-driven letter-level timing, syntactically structured pause placement, and content-correlated burst patterns---a multi-level coherence constraint that no current attack addresses. Banerjee et al.~\cite{Banerjee2014} further showed that cognitive-affective states (including deception) leak into typing patterns in ways that parallel speech prosody: deceptive writers exhibited irregular pause distributions and atypical editing rhythms compared to truthful writers, analogous to the disfluency and pitch variation observed in deceptive speech. This raises the possibility that an adversary attempting to fake cognitive signatures may introduce artifacts of deliberate control---unnaturally regular pause placement, suppressed revision activity, or anomalous burst uniformity---that are themselves detectable as markers of performance rather than composition.

\paragraph{The information-theoretic asymmetry.} Motor features occupy a low-dimensional space: the Aalto 136M dataset yields a population entropy of approximately 4.12 bits for IKI distributions, and individual motor profiles can be characterized by fewer than 20 distributional parameters. Cognitive features are entangled with the semantic content being produced: the correct pause distribution depends on \emph{what} is being written, not just \emph{how fast}. Forging the cognitive channel requires reproducing the correct timing conditioned on specific content---which in turn requires the adversary to replicate the writer's cognitive process in real time. The non-identifiability result from Condrey~\cite{Condrey2026TrustInversion}---that mutual information between motor features and content origin is zero for copy-type attacks---does not extend to cognitive features precisely because CLC is bound to semantic content. Mehta et al.~\cite{Mehta2025} provide empirical support: machine learning models distinguishing genuine composition from AI-text transcription achieved F1 scores exceeding 97\%, while human evaluators performed at chance ($\sim$45\% F1), suggesting that the cognitive signal is both real and difficult to consciously reproduce.

\paragraph{Acknowledged limitations and adversarial cost.} A sufficiently motivated adversary who pre-composes text, memorizes it, and practices typing it with content-appropriate pauses could plausibly produce a positive CLC. However, this transforms the adversary's task from trivial copy-paste (seconds) to an elaborate rehearsed performance. We estimate the minimum attack cost as follows: the adversary must (1)~generate or obtain the text, (2)~analyze its complexity profile to identify where planning pauses should occur, (3)~memorize the text and its pause map, and (4)~execute the performance under time pressure while maintaining burst-level and revision-level consistency. Steps 2--4 require approximately 30--60 minutes of preparation per 1,500-word document, based on the time needed to analyze complexity at ${\sim}150$ clause boundaries, memorize the text, and rehearse content-appropriate pause placement. This preparation time is comparable to composing the document from scratch, which collapses the economic incentive for the attack. Under multi-session consistency checks, the adversary must additionally maintain a stable CLC profile across documents and sessions, compounding the rehearsal cost. A single-session evasion may be feasible with moderate effort, but each additional session multiplies the constraints: the adversary's CLC distribution, burst-length statistics, and revision frequency must remain consistent with a plausible individual baseline while varying naturally across topics and writing conditions. Because these parameters are correlated in genuine writers but independently controlled in a rehearsed performance, statistical tests for inter-session consistency (e.g., comparing within-writer CLC variance to population norms) should flag fabricated profiles with increasing reliability as the number of observed sessions grows. We recommend layered verification combining CLC with revision-pattern analysis and session-level consistency checks to address these residual risks.

\section{INTERFACE DESIGN AND ETHICAL DEPLOYMENT}
\label{sec:interface-design}

Deploying a behavioral biometric system within writing interfaces requires careful attention to transparency, consent, and fairness. We organize our design principles around four concerns.

\subsection{Layered Transparency}

The system must clearly communicate what data is collected (keystroke timestamps), what is not collected (key identities, screen content), and how the data is used (aggregate CLC computation, not keystroke replay). We propose a three-layer transparency model: (1)~a persistent but unobtrusive status indicator showing that timing collection is active; (2)~a detail panel accessible on demand showing the current CLC estimate and data retention status; and (3)~a pre-deployment disclosure document explaining the verification methodology in plain language.

\subsection{Consent and User Control}

Writers must opt in to verification. The system should provide controls to pause collection, delete stored timing data, and export a copy of one's own evidence record. In educational deployments, consent must be genuinely voluntary---students should not face grade penalties for declining, and alternative verification pathways (e.g., oral defense) should be available.

\subsection{Contextual Integrity in Educational Settings}

Collecting behavioral biometrics from students in learning management systems raises power-asymmetry concerns that go beyond standard informed consent. Nissenbaum's~\cite{Nissenbaum2004} contextual integrity framework requires that information flows conform to the norms of the context in which they occur. Students typing in a text editor expect that their \emph{words} will be evaluated, not their \emph{pauses}. Verification systems must be framed transparently as tools that protect the student's claim to authorship---not as surveillance instruments deployed by the institution. Data retention should be minimized: once a verification decision is rendered, raw timing data should be deleted, retaining only the aggregate CLC score and a cryptographic commitment.

\subsection{Biometric Parity}

Users with motor impairments, repetitive strain injuries, or atypical input methods (switch access, eye-gaze typing) may produce timing distributions that confound the cognitive-motor separation assumed by CLC. Users with visual impairments who rely on screen readers, braille displays, or refreshable input devices face a distinct challenge: their keypress timing is mediated by auditory or tactile feedback loops that introduce latencies unrelated to cognitive load. The framework should include a calibration phase that establishes a personal baseline, and a biometric parity mode that adjusts CLC thresholds relative to the individual's own motor profile rather than population norms. Where keystroke verification is infeasible (e.g., voice-to-text users, screen-reader-dependent workflows), alternative process attestation methods should be available, and the system must not penalize users for whom the default modality is inaccessible.

\subsection{Position Within the Process Attestation Program}

The cognitive signatures framework presented here forms the behavioral evidence domain of a broader process attestation architecture. Companion work~\cite{Condrey2026TrustInversion} formalizes the trust inversion threat model. The CLC metric defined in this paper serves as the primary behavioral discriminator in a pipeline that combines behavioral, temporal, and content evidence domains. Integration with complementary evidence streams---including revision-pattern analysis and session-level consistency checks---is the subject of ongoing work.

\section{CONCLUSION}
\label{sec:conclusion}

Authorship is not a property of text. It is a property of the human process that created it. We have characterized the cognitive signatures embedded in typing behavior during genuine composition, defined the Cognitive Load Correlation as a discriminative measure, and presented a non-intrusive framework for collecting process evidence within existing writing interfaces. Our adversarial analysis shows that cognitive signatures resist the timing-forgery attacks that defeat motor-level keystroke authentication, because the cognitive channel is entangled with semantic content in ways that cannot be reproduced without replicating the writing process itself.

Reframing authorship verification as a human-computer interaction problem moves beyond the information-theoretic limits of output-level detection. Where output-based detectors degrade as model distributions converge toward human distributions, process-based verification exploits a signal that AI-generated text cannot produce: the temporal trace of a mind composing in real time.

Several limitations bound the present contribution. Our evaluation on the KLiCKe corpus (AUC 0.83 in LOWO-CV across 4,971 writers) demonstrates measurable discrimination, but the population scope is narrow: the corpus skews toward university-educated adults on desktop keyboards. Cross-population validation---particularly for non-native writers, mobile-keyboard users, and individuals with motor impairments---remains essential before deployment. A controlled within-subjects study comparing composition and transcription with concurrent content-complexity annotations would strengthen the causal claims.

Future work should prioritize four directions: (1)~a within-subjects user study collecting both composition and transcription keystroke logs with concurrent content-complexity annotations; (2)~longitudinal evaluation of CLC stability across writing sessions, topics, and fatigue states; (3)~integration of CLC with complementary signals such as revision-pattern analysis and session-level consistency to build layered verification systems; and (4)~cross-population validation covering children and elderly adults, non-native writers, users of non-Latin writing systems, and individuals relying on accessibility input modalities---populations for which the current framework provides no empirical coverage.

The keyboard has always been a window into the mind. It is time we started reading the evidence it already provides.

\section*{REFERENCES}
\begingroup
\renewcommand{\section}[2]{}
\bibliographystyle{IEEEtran}
\bibliography{refs}
\endgroup

\end{document}